\documentclass{article}
\usepackage{spconf,amsmath,graphicx}
\usepackage{booktabs}
\usepackage{tabularx}
\usepackage{url}
\usepackage{float}
\usepackage{algpseudocode}
\usepackage[dvipsnames]{xcolor}

\title{Classification of COVID-19 X-ray Images Using a Combination of 
Deep and Handcrafted Features}
\name{Weihan Zhang$^{\star}$ \qquad Bryan Pogorelsky$^{\star}$ \qquad Mark Loveland$^{\dagger}$ \qquad Trevor Wolf$^{\star}$}
  
\address{$^{\star}$ Dept. of Aerospace Engineering and Engineering Mechanics, The University of Texas at Austin\\
      $^{\dagger}$ Oden Institute for Computational Engineering and Sciences, The University of Texas at Austin }

\begin{document}
%
\maketitle

\begin{abstract}
Coronavirus Disease 2019 (COVID-19) demonstrated the need for accurate and fast diagnosis methods for emergent viral diseases. Soon after the emergence of COVID-19, medical practitioners used X-ray and computed tomography (CT) images of patients' lungs to detect COVID-19. Machine learning methods are capable of improving the identification accuracy of COVID-19 in X-ray and CT images, delivering near real-time results, while alleviating the burden on medical practitioners. In this work, we demonstrate the efficacy of a support vector machine (SVM) classifier, trained with a combination of deep convolutional and handcrafted features extracted from X-ray chest scans. We use this combination of features to discriminate between healthy, common pneumonia, and COVID-19 patients. The performance of the combined feature approach is compared with a standard convolutional neural network (CNN) and the SVM trained with handcrafted features. We find that combining the features in our novel framework improves the performance of the classification task compared to the independent application of convolutional and handcrafted features. Specifically, we achieve an accuracy of 0.988 in the classification task with our combined approach compared to 0.963 and 0.983 accuracy for the handcrafted features with SVM and CNN respectively.

\end{abstract}
\section{Introduction}
\label{sec:intro}

Coronavirus disease 2019 (COVID-19) is an infectious disease caused by Severe Acute Respiratory Syndrome Coronavirus 2 (SARS-CoV-2). Since the emergence in Wuhan, China in December 2019, it has spread worldwide and has caused a severe pandemic. The COVID-19 infection causes mild symptoms in the initial stage, but may lead to severe acute symptoms like multi-organ failure and systemic inflammatory response syndrome \cite{wu2020characteristics,goyal2020clinical}. As of December 2020, there have been more than 1.8 million COVID-19 related deaths around the world and daily new cases of the disease are still rising. Currently, reverse transcription polymerase chain reaction test (RT-PCR) is the most accurate diagnostic test. However, it requires specialized materials, equipment, personnel, and takes at least 24 hours to obtain a result. It may also require a second RT-PCR or a different test to confirm the diagnosis. Therefore, radiological imaging techniques like X-ray and CT-scan can serve as a complement to improve diagnosis accuracy \cite{ai2020correlation}. 

In recent years, machine learning has been used extensively for automatic disease diagnosis in the healthcare sector \cite{sajda2006machine,miotto2018deep}. Various standard supervised learning algorithms such as logistic regression, random forests, and support vector machines (SVM) have been applied in detecting COVID-19 in X-ray and CT images of patients' lungs \cite{li2020clinical,tang2020severity,barstugan2020coronavirus,sethy2020detection}. The convolutional neural network (CNN) is a deep learning algorithm that can extract features from images through a combination of convolutional, pooling, and fully connected layers. It has been used extensively for image recognition, classification, and object detection. Recent works \cite{apostolopoulos2020COVID,asnaoui2020automated,ozturk2020automated,mahmud2020covxnet,hall2020finding} show that it can also provide accurate results in detecting COVID-19 in images. These recent works present some insightful thoughts and valuable opinions. However, the lack of publicly available image databases and the limited amount of patient data are inevitable challenges for training a CNN.

In this study, we propose a fusion model that classifies X-ray images from a combination of handcrafted features and CNN deep features. The model is trained and tested on a large dataset with 1,143 COVID-19 cases, 2,000 normal cases and 2,000 other pneumonia cases collected from \cite{kermany2018labeled,kaggle}. The feature fusion classifier has been shown as an effective way of boosting the performance of CNN models in face recognition \cite{nguyen2018combining} and biomedical image classifications \cite{nanni2018bioimage,srivastava2020classification}. Handcrafted and deep features extract different information from the same input image, so the fusion of these two systems has the potential to outperform the standard approaches \cite{nanni2017handcrafted}. Our key interest is whether a fusion model can also surpass the standard CNN and SVM for COVID-19 detection. The paper is organized as follows: The methodology and feature extraction techniques are presented in Section 2, the comparative classification performances are given in Section 3, and the final conclusions are made in Section 4.

\section{Methodology}
\label{sec:Methodology}
\begin{figure}[htb]
\begin{minipage}[t]{.32\linewidth}
  \centering
  \centerline{\includegraphics[width=2.7cm]{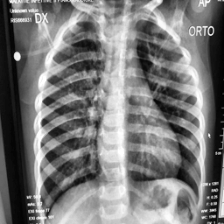}}
  \centerline{(a) COVID-19}
\end{minipage}
\begin{minipage}[t]{.32\linewidth}
  \centering
  \centerline{\includegraphics[width=2.7cm]{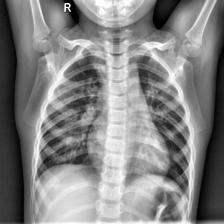}}
  \centerline{(b) Normal}
\end{minipage}
\begin{minipage}[t]{0.32\linewidth}
  \centering
  \centerline{\includegraphics[width=2.7cm]{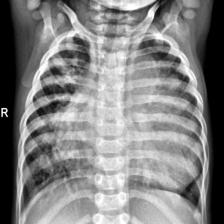}}
  \centerline{(c) Pneumonia}\medskip
\end{minipage}
\caption{Sample images after preprocessing}
\label{example}
\end{figure}

\begin{figure*}[htb!]
  \centering
  \includegraphics[width=0.8\linewidth]{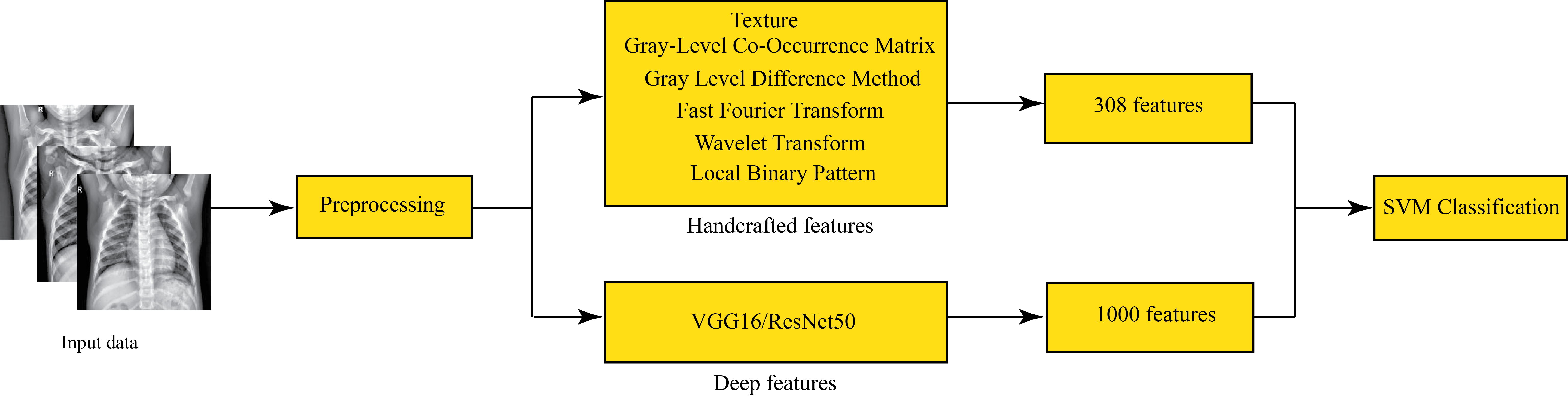}
  \caption{Methodology}
  \label{Methodology}
\end{figure*}

The proposed COVID-19 classifier is trained and tested on a collective dataset with 5,143 X-ray images categorized into three cases: COVID-19, Normal and Pneumonia. All the images are resized to 224 $\times$ 224 pixels and the local contrast is enhanced by an adaptive histogram equalization algorithm during the preprocessing stage. Several preprocessed example images are show in Figure \ref{example}. Both handcrafted features and VGG16/ResNet50 deep features are extracted from the dataset, then combined and fed into an SVM classifier. The entire process is shown in Figure \ref{Methodology}.

\subsection{Handcrafted Features}
\label{sec:Handcrafted}
Handcrafted features seek to characterize each image by computing properties using the information directly present in each image. These handcrafted features are computed for each image and used as input into the SVM. There are 308 features computed on each image by evaluating 14 different statistical measures on the output of each image under different transformations. The transformations can be categorized into six groups: Texture, Gray-Level Co-Occurrence Matrix (GLCM), Gray Level Difference Method (GLDM), Fast Fourier Transform, Wavelet transform, and Local Binary Pattern.  The features are computed by applying the following same 14 statistical measures on the outputs from the aforementioned six transformations: area, mean, standard deviation, skewness, kurtosis, energy, entropy, maximum, mean absolute deviation, median, minimum, range, root mean square, and uniformity as used in a COVID-19 image classifier that used handcrafted features only \cite{Khuzani2020}. Of the aforementioned 14 measures, the following 10 are all calculated using the standard definitions: Mean, standard deviation, maximum, minimum, median, range, root mean square, skewness, mean absolute deviation, and kurtosis. Energy was calculated using the following definition: 
\begin{equation}
    Energy :=\sum_{i=1}^{length(p)} p_i^2
\end{equation}
where $p_i$ is the $i^{th}$ value from the output vector of a transformation. Area here is defined as the sum of all of the components of the output vector. Entropy is calculated by first taking the frequency of each unique intensity via the numpy function unique() and then normalizing that vector. From there the entropy is directly calculated by taking the elementwise sum of that normalized vector times the base 2 log of itself. Uniformity is also calculated from this normalized vector. For clarity, the pseudo-code is reproduced below:
	\begin{algorithmic}
	\Require $p$ (vector of output from a transformation)\\
	$value, counts = unique(p,returncounts=True)$\\
	$counts = counts / (\sum_i^{length(p)} counts_i)$ \\
	$entropy = -\sum_i^{length(p)}counts_i*log_2(counts_i)$ \\
	$uniformity= \sum_i^{length(p)} counts_i^2$ \\
    \end{algorithmic}

\begin{itemize}
\item \textbf{Texture}
The texture features are calculated by considering each input image as a single row vector and then calculating each of the above metrics on the vector. For example the texture feature corresponding to the mean is simply the sum of all of the pixel values (integer from 0 to 255) divided by the number of pixels in the image. This results in a total of 14 features computed.  

\item \textbf{GLCM}
The GLCM transform characterizes an image by creating a histogram of co-occurring greyscale values at a given offset and direction over an image \cite{Haralick1973}. In this specific implementation of GLCM, features are determined by applying the greycomatrix() function from the skimage library directly on each image with an offset of 1 and in four different directions (0, $\pi$/4, $\pi$/2, 3$\pi$/4). This function returns a 4-D array corresponding to each direction. Each dimension is evaluated on the 14 statistical measures as before, resulting in a total of 56 features.

\item \textbf{GLDM}
GLDM is a method that characterizes an image by creating a distribution of the absolute differences of pixel intensity to the pixel intensity of surrounding pixels at a given distance and direction \cite{Conners1980}. In this implementation, GLDM is computed in four directions (0, $\pi$/2, $pi$, 3$\pi$/2) with a distance of 10 pixels. Each of the four directions gives an output vector and the 14 statistical measures are computed on each output resulting in 56 features

\item \textbf{FFT}
The FFT features are evaluated on each image by transforming the image via a Fast Fourier Transform. Each image is input into the numpy fft.fft2() function resulting in a vector of output values that are then put into the numpy fft.fftshift() function. Next, the numpy floor() function is used to convert to a vector of integers which is the final output that is used to compute the 14 statistical measures which are the FFT features.

\item \textbf{Wavelet}
The wavelet features are computed by applying the pywt package's dwt2() function directly on each image \cite{Lee2019}. The output of this function gives 4 different arrays and the 14 statistical measures are computed on each array resulting in 56 features. The first array from the dwt2() output is then put back into the dwt2() function as input, resulting in another 4 matrices. Again, these 4 matrices are used to compute 14 statistical measures each for another 56 features resulting in 112 features in total from the wavelet transforms.

\item \textbf{LBP}
LBP works by looking at points surrounding each pixel within a given distance and tests whether the points are greater than or less than the central point resulting in a binary output \cite{LBP1996}. In this implementation, scikit-image's local\_binary\_pattern function is used to compute the LBP outputs with distances of 2,3,5, and 7. The resulting four LBPs are then used to compute the 14 statistical measures resulting in 56 features.

\end{itemize}

\subsection{Deep Features}
\label{sec:Deep}
Deep features are extracted from two CNN models, VGG16 \cite{simonyan2014very} and ResNet50 \cite{he2016deep}. More specifically, only the feature extraction layers of the model are utilized which are positioned prior to dense layers meant for the classification task. The model weights are pre-trained on the ImageNet dataset \cite{russakovsky2015imagenet} which contains millions of images belonging to 1,000 classes. An important note is that no fine tuning is done to the models, meaning that the model weights are fixed and no further training is done.

\begin{figure}[htb]
  \centering
  \includegraphics[width=0.47\textwidth]{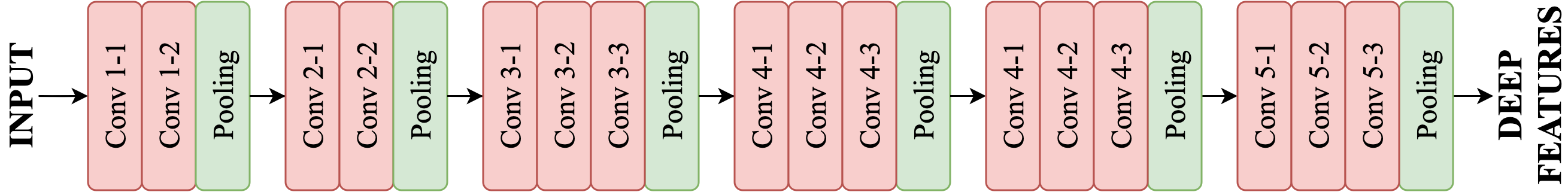}
  \caption{VGG16 feature extraction layers}
  \label{VGG16_arch}
\end{figure}

The VGG16 CNN architecture contains 16 layers with trainable weights (with 5 being dense layers that are not used for feature extraction) consisting of 5 blocks that include convolutional and pooling layers which can be seen in Figure \ref{VGG16_arch}. The input of the model accepts RGB images of size 224 $\times$ 224 $\times$ 3 pixels. To maintain compatibility with the model, the grayscale the X-ray images are converted to to have three color channels by simply duplicating the pixel values and having each color channel be identical. Additionally, each image is zero-centered with respect to the ImageNet dataset without scaling. For each X-ray image the resulting feature output is of dimension 7 $\times$ 7 $\times$ 512 with subsequent flattening producing a vector containing 25,088 features.

\begin{figure}[htb]
  \centering
  \includegraphics[width=0.2\textwidth]{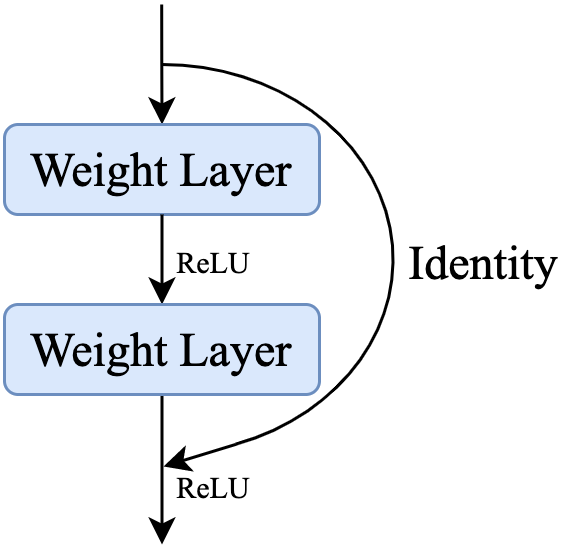}
  \caption{Residual block}
  \label{res_block}
\end{figure}

As opposed to CNN architectures such as VGG, ResNets can have more layer depth with increasing accuracy while at the same time having less overall complexity. This is achieved by utilizing shortcut connections allowing residual mapping that may skip one or more layers and performing identity mapping which can alleviate the problem of vanishing gradients. A residual block of this type is shown in Figure \ref{res_block}. The ResNet50 model contains 50 layers with trainable weights (of which a single dense layer is not used for feature extraction). As with VGG16 the input of the model accepts RGB images of size 224 $\times$ 224 $\times$ 3 pixels. Again, the grayscale X-ray images are converted to duplicated three channel RGB before being zero-centered with respect to the ImageNet dataset without scaling. Each X-ray image results in a feature output of dimension 7 $\times$ 7 $\times$ 2048 and is flattened into a vector of 100,352 features.

After features are extracted from the models, kernel principal component analysis (PCA) is applied to reduce the dimensionality of the deep features. The number of components after the transformation is selected to be 1,000 as this number of features is near the order of magnitude as the number of handcrafted features that are extracted.

\subsection{Classifier}
A linear SVM using one-vs-all approach is applied to classify the combined features. Despite the fact that most deep learning models employ the softmax activation function for classification task, it was shown that SVM works better on several standard datasets like MNIST,CIFAR-10,and the ICML 2013 Representation Learning Workshop’s face expression recognition challenge \cite{tang2013deep}. 
\label{sec:SVM}

\section{Results and Discussions}
\label{sec:Result}
To evaluate the performance of the method outlined above, it was important to compare the performance of combined deep features and handcrafted features in an SVM classifier with baseline individual CNNs in addition to solely using the handcrafted features in an SVM. 

Both the VGG and ResNet CNNs were evaluated again with the feature extraction layers frozen with pre-trained ImageNet weights. Two layers were added to the models, a 1,000 neuron dense layer with a rectified linear activation function and a three neuron output layer with a softmax activation function. The objective of the addition of the layers is to allow the classification of three classes to be possible in addition to increasing the number of trainable parameters as the feature extraction layers are frozen. Additionally, during training both models use categorical cross-entropy while employing the Adam optimizer \cite{kingma2014adam} with a learning rate of 0.005.

A parametric study was performed on the handcrafted features to evaluate which configuration created most accurate results as inputs into the SVM. Results in Table \ref{tab:handcrafted} show that by itself, the Wavelet features resulted in the highest classification accuracy followed by GLDM and GLCM. The lowest performing feature group was the texture features with an accuracy of 0.762. It was found that inputting all features (308) into the SVM resulted in the highest accuracy and F-1 Score. A 95\% confidence interval is given for all values in Table \ref{tab:handcrafted}.

\begin{figure}[H]
\centering
\begin{minipage}[t]{0.49\linewidth}
  \centering
  \includegraphics[width=4cm]{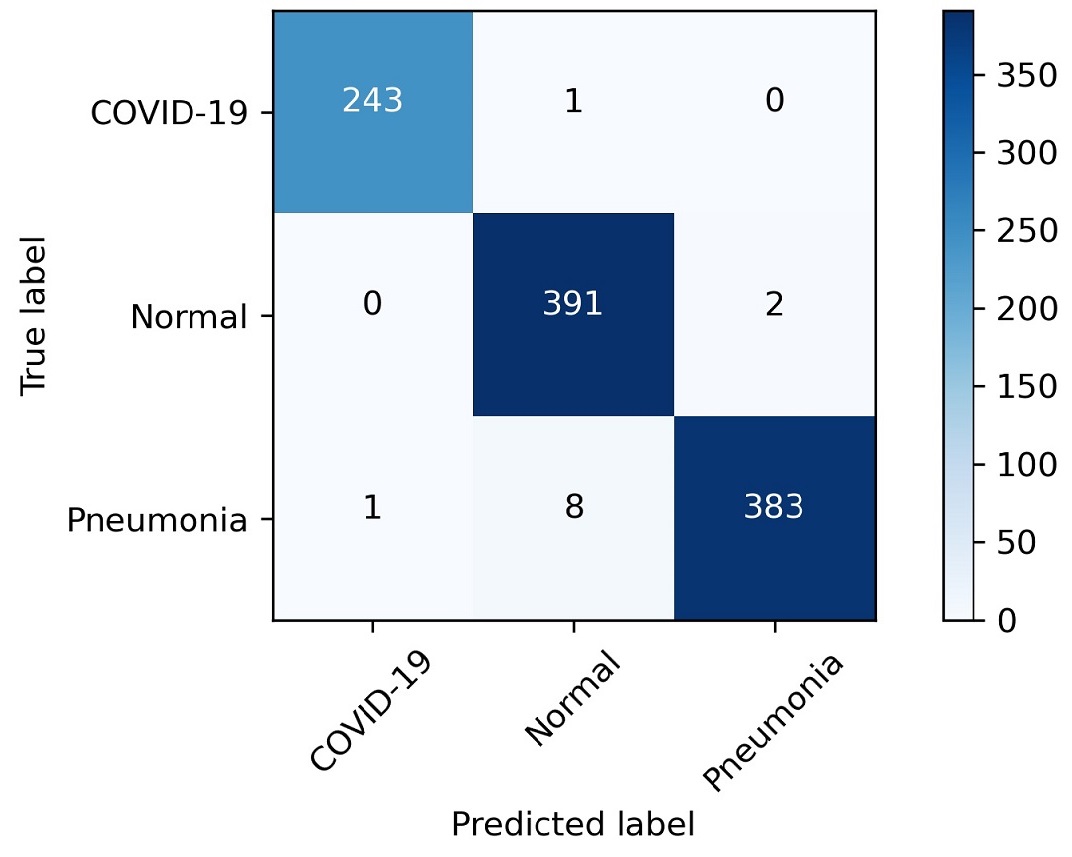}
  \centerline{(a) VGG16 DF + HF}
\end{minipage}
\begin{minipage}[t]{0.49\linewidth}
  \centering
  \includegraphics[width=4cm]{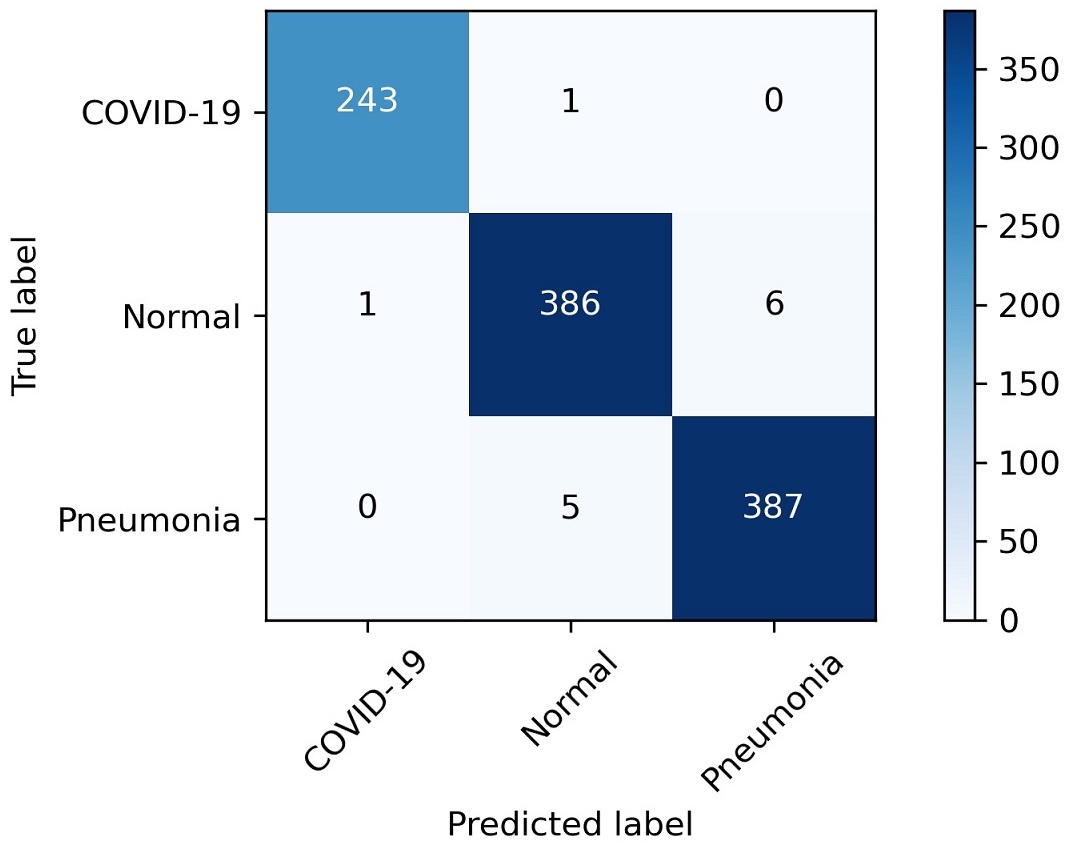}
  \centerline{(b) ResNet50 DF + HF}
\end{minipage}
\caption{Confusion matrices}
\label{CM}
\end{figure}

For each classification model outlined above, the dataset of 5,143 was divided into the same train and test subsets with an 80/20 split. This resulted in 4,114 training images and 1,029 test images. The results of each classification model can be seen in Table \ref{tab:results}. All the metrics listed in the table are unweighted averages of the statistics of each class with a 95\% confidence interval. From these results its is clear that all models that incorporate deep features clearly performed better than the SVM that only uses handcrafted features. The two models utilizing both deep features and handcrafted featured with an SVM classifier slightly outperform the conventional VGG16 and ResNet50 CNNs. Additionally, the confusion matrices of the combined deep features and handcrafted features SVM models are seen in Figure \ref{CM}. Both combined feature models achieve the same low false negative and false positive rates of 0.41\% and 0.13\% respectively.

\begin{table}[htb]
\small
\centering
\caption{Performance of X-ray image classification using SVM with handcrafted features only}
\label{tab:handcrafted}
\begin{tabular}{@{}ccc@{}}
\toprule
\textit{\textbf{Handcrafted Features}} & \textit{\textbf{Accuracy}} &  \textit{\textbf{F1-Score}} \\ \midrule
Texture                          &    0.762 $\pm$ 0.026      &   0.771 $\pm$ 0.026    \\
GLCM  &    0.896 $\pm$ 0.019      &   0.880 $\pm$ 0.020    \\
GLDM     &    0.900 $\pm$ 0.018       &   0.894 $\pm$ 0.019    \\
FFT           &    0.818 $\pm$ 0.024      &   0.809 $\pm$ 0.024   \\
Wavelet                          &    0.940 $\pm$ 0.015      &   0.934 $\pm$ 0.015 \\
LBP             &    0.874 $\pm$ 0.020       &   0.880 $\pm$ 0.020    \\
All Features Combined            &    0.963 $\pm$ 0.012     &   0.957 $\pm$ 0.012   \\\bottomrule
\end{tabular}
\end{table}

\begin{table}[htb]
\small
\centering
\caption{Performance of X-ray image classification models}
\label{tab:results}
\begin{tabular}{@{}ccc@{}}
\toprule
\textit{\textbf{Classification Model}} & \textit{\textbf{Accuracy}} &  \textit{\textbf{F1-Score}} \\ \midrule
Handcrafted Features (SVM)          &   0.963 $\pm$ 0.012    &    0.957 $\pm$ 0.012     \\
VGG16                               & 0.982 $\pm$ 0.008 & 0.983 $\pm$ 0.008\\
ResNet50                            &    0.983 $\pm$ 0.008   &   0.984 $\pm$ 0.008   \\
VGG16 DF + HF (SVM)    &    0.988 $\pm$ 0.007      &   0.989 $\pm$ 0.006  \\
ResNet50 DF + HF (SVM)& 0.987 $\pm$ 0.007 & 0.988 $\pm$ 0.007 \\ \bottomrule
\end{tabular}
\end{table}

\setlength{\parindent}{0em}
\setlength{\parskip}{-0.2em}

\section{Conclusion}
\label{sec:Conclusion}
This work demonstrated the use of a combined handcrafted and deep feature approach for classifying COVID-19, pneumonia, and healthy patients in radiological images. This new approach was compared to 7 handcrafted feature classifiers and two CNN architectures. With respect to all performance metrics, the combination of deep features and handcrafted features surpassed that of handcrafted features or deep features alone. Notably, the proposed architecture achieved an accuracy of 0.988 by combining VGG16 deep features and handcrafted features. The next best accuracy of an approach without combining deep and handcrafted features was 0.983 for ResNet50.

\small
\bibliographystyle{IEEEbib}
\bibliography{ref}

\end{document}